\documentclass[preprint]{aastex63}

\received{April 8 2021}
\revised{May 6 2021}
\accepted{May 7 2021}
\submitjournal{ApJ}

\shorttitle{Kepler RGS analysis}
\shortauthors{Kasuga et al.}

\begin{document}

\title{Spatially Resolved RGS Analysis of Kepler's Supernova Remnant}

\correspondingauthor{Tomoaki Kasuga}
\email{tomoaki.kasuga@phys.s.u-tokyo.ac.jp}

\author[0000-0001-5241-9768]{Tomoaki Kasuga}
\affiliation{Department of Physics, Graduate school of Science, The University of Tokyo, 7-3-1 Hongo, Bunkyo-ku, Tokyo 113-0033, Japan}

\author[0000-0002-4708-4219]{Jacco Vink}
\affiliation{Anton Pannekoek Institute for Astronomy, University of Amsterdam, Science Park 904, 1098 XH Amsterdam, The Netherlands}
\affiliation{GRAPPA, University of Amsterdam, PO Box 94249, 1090 Amsterdam, The Netherlands}
\affiliation{SRON, Netherlands Institute for Space Research, Sorbonnelaan 2, 3584 Utrecht, The Netherlands}

\author[0000-0002-1104-7205]{Satoru Katsuda}
\affiliation{Graduate School of Science and Engineering, Saitama University, 255 Shimo-Ohkubo, Sakura, Saitama 338-8570, Japan}

\author[0000-0003-1518-2188]{Hiroyuki Uchida}
\affiliation{Department of Physics, Kyoto University, Kitashirakawa Oiwake-cho, Sakyo, Kyoto, Kyoto 606-8502, Japan}

\author[0000-0003-0890-4920]{Aya Bamba}
\affiliation{Department of Physics, Graduate school of Science, The University of Tokyo, 7-3-1 Hongo, Bunkyo-ku, Tokyo 113-0033, Japan}
\affiliation{Research Center for the Early Universe, School of Science, The University of Tokyo, 7-3-1 Hongo, Bunkyo-ku, Tokyo 113-0033, Japan}

\author[0000-0001-9267-1693]{Toshiki Sato}
\affiliation{Department of Physics, Rikkyo University, 3-34-1 Nishi Ikebukuro, Toshima-ku, Tokyo 171-8501, Japan}
\affiliation{RIKEN, 2-1 Hirosawa, Wako, Saitama 351-0198, Japan}

\author[0000-0002-8816-6800]{John P. Hughes}
\affiliation{Department of Physics and Astronomy, Rutgers University, 136 Frelinghuysen Road, Piscataway, NJ 08854-8019, USA}

\begin{abstract}
The distribution and kinematics of the circumstellar medium (CSM) around a supernova remnant (SNR) tell us useful information about the explosion of its natal supernova (SN). 
Kepler’s SNR, the remnant of SN1604, is widely regarded to be of Type Ia origin. 
Its shock is moving through a dense, asymmetric CSM. 
The presence of this dense gas suggests that its parent progenitor system consisted of a white dwarf and an asymptotic giant branch (AGB) star. 
In this paper, we analyze a new and long observation with the reflection grating spectrometers (RGS) on board the {\it XMM-Newton} satellite, spatially resolving the remnant emission in the cross-dispersion direction. 
We find that the CSM component is blue-shifted with velocities in the general range 0--500~km s${}^{-1}$. 
We also derive information on the central bar structure and find that the northwest half is blue-shifted, while the southeast half is red-shifted. 
Our result is consistent with a picture proposed by previous studies, in which a ``runaway'' AGB star moved to the north-northwest and toward us in the line of sight, although it is acceptable for both single-degenerate and core-degenerate scenarios for the progenitor system.
\end{abstract}

\keywords{ISM: supernova remnants, X-rays: individual (Kepler's SNR), supernovae: individual (SN1604), circumstellar matter}

\section{Introduction} \label{sec:intro}

Kepler's supernova remnant (SNR) is the remnant of SN 1604 \citep{Baade43}, the last naked-eye Galactic supernova and named after
Johannes Kepler, who wrote a monograph on the appearance and the brightness evolution of this new star.
Kepler's SNR  is one of the best studied SNRs across the electromagnetic spectrum \citep[see][for a review]{Vink17}. 
Like the similarly aged SN 1572/Tycho's SNR  \citep[see][for a review]{Decourchelle17}, Kepler's SNR is identified as the remnant of a Type Ia supernova  \citep{Kinugasa99,Reynolds07,Badenes06}. 
Unlike the remnant of SN 1572, Kepler's SNR is not spherically symmetric \citep{Lopez11}.
The probable reason is that the progenitor system for SN 1604 was a runaway system, and the SNR is interacting with a bow-shock shaped shell concentrated toward the northwest, created by the interaction of a stellar wind from the moving progenitor system with the local, tenuous interstellar medium.
The evidence for this was first reported by  \citet{Bergh77,Bandiera91}, who found that the optical emission from dense knots has an overall proper motion of $\sim200$~km\ s$^{-1}$ out of the Galactic plane, whereas narrow-line H$\alpha$ emission shows blueshifts of around 180~km\ s$^{-1}$
 \citep{Blair91}. \par
The SNR distance was long uncertain, but recent estimates based on the kinematics and the historical light curve suggest a distance of $\approx 5\pm 1$~kpc \citep{Vink17,Ruiz17,Sankrit16}, implying a height above the Galactic plane of $\approx 594d_5$~pc --- with $d_5$ the distance in units of 5~kpc.
The velocity of the progenitor system resulted in a density that is highest in the northwestern region, where the SNR blast wave has indeed slowed down considerably \citep{Vink08,Katsuda08}. \par
The presence of nitrogen-rich material was initially interpreted as due to the wind of a massive star \citep{Bandiera87,Borkowski92}. 
But the over-abundance of iron-rich ejecta, and lack of oxygen-rich ejecta as seen in the X-ray band \citep{Kinugasa99,Reynolds07} clearly suggest a Type Ia origin. 
There are two main scenarios for the progenitor systems for Type Ia supernova explosions. 
One is the single degenerate (SD) scenario, for which the binary progenitor system consists of a C+O white dwarf (WD) and a non-degenerate star, for example an asymptotic giant branch (AGB) star \citep{Whelan73,Nomoto82,Han04}. 
Due to a mass transfer, the mass of the WD is increased to near the Chandrasekhar mass ($\sim1.4$~$M_{\odot}$), which results in high density core, initiating a thermonuclear explosion. 
The other scenario is the double degenerate (DD) scenario. According to this scenario the progenitor system consists of a WD-WD binary.
The explosion is initiated when the WDs merge due to orbital energy loss caused by gravitational wave emission \citep{Iben84,Webbink84}.
A third scenario that is sometimes discussed is the so-called core degenerate (CD) scenario, which is somewhat  intermediate between the SD and DD scenarios.
It involves the merger of a WD and an AGB core in a WD-AGB binary system \citep{Livio03,Ilkov12,Kashi11}. \par
The presence of a dense shell at $\sim 600$~pc above the Galactic plane, which is not only rich in nitrogen, but also silicate-rich dust grains \citep{Douvion01,Williams12}, suggests that the SN 1604 progenitor system consisted of a  ``runaway" system containing a WD and an AGB star. 
This model was explored by 2D \citep{Velazquez06,Chiotellis12,Burkey13} and a 3D \citep{Toledo-Roy14} hydrodynamical simulations. 
Silicate-rich dust is typically produced by relatively massive AGB stars \citep[$\gtrsim 4$~$M_\odot$, e.g.][]{Karakas10}.
However, the core of the AGB donor star is expected to survive the Type Ia explosion, which is at odds with the lack of bright stars near the explosion centre in Kepler's SNR \citep{Kerzendorf14,Ruiz18}. 
This puts Kepler's SNR in a similar situation as other Type Ia SNRs: Tycho's SNR, SNR 0509-675, and SN 1006 \citep{Kerzendorf13,Schaefer12,Gonzalez12}, namely favouring a DD-scenario progenitor, if it were not for the CSM structure and composition. 
More recently, \citet{Chiotellis20} simulated Kepler's SNR assuming a CD scenario, in which the short-lived double degenerate system merges while still situated within the planetary nebula created by the AGB star.
This scenario also explains the ``ears'' in the north-west and south-east of the remnant, but also retains the bowshock shaped region due to proper motion of the progenitor system \citep{Tsebrenko13,Chiotellis21}. \par
For a better understanding of the supernova explosion and interaction with the CSM, we need to explore in more detail the kinematics.
In X-rays, kinematic studies have concentrated either on the proper motions \citep{Vink08,Katsuda08}, rather than on the line of sight motions\citep{Kasuga18}.
Recently, a study was published combining the proper motions of individual bright X-ray knots, with Doppler measurements obtained from {\it Chandra} ACIS-S data \citep{Sato17b} and HETGS data \citep{Millard20}. 
However, the dispersion angle of the HETGS does not allow to study the Doppler shifts and Doppler broadening from the SNR in its entirety. \par
Here we explore the data of Kepler's SNR obtained with the  Reflection Grating Spectrometers (RGS) \citep{Herder01} on board the {\it XMM-Newton} satellite \citep{Jansen01}. 
The spectra are suitable for decomposing detailed line features in the lower X-ray energy band, including CSM emission lines. 
A detailed analysis of the RGS data for Kepler's SNR was carried out by \citet{Katsuda15}. 
Here we follow-up on this study, by using deeper RGS observations.
Apart from obtaining a better view of the net Doppler shifts of both the ejecta and CSM components, the Doppler broadening provides additional information on both the kinematics and thermal + turbulent Doppler broadening, which yields constraints on the ion temperature. \par
This paper is structured as follows. 
Section \ref{sec:obs} describes the observation information and data reduction process. 
Section \ref{sec:analysis} explains the details of our analysis including the process of dividing regions and fitting models. 
Finally, section \ref{sec:discussion} discusses our results concerning mainly to the kinematics of CSM.

\section{Observations} \label{sec:obs}

\begin{deluxetable*}{cccccc}[!ht]
\tablecaption{RGS observation table\label{tab:rgs_obs}}
\tablewidth{0pt}
\tablehead{
\colhead{Target} & \colhead{Obs. ID} & \colhead{Time} & \colhead{Position} & \colhead{Roll angle} & \colhead{Exposure} \\
\colhead{} & \colhead{} & \colhead{} & \colhead{(RA,dec)} & \colhead{(deg)} & \colhead{(ksec)}
}
\startdata
Kepler's SNR & 0084100101 & Mar. 2001 & (262.67499, -21.491667) & 94.71 & 31.7 \\
Kepler's SNR & 0853790101 & Mar. 2020 & (262.67499, -21.491667) & 94.68 & 37.0 \\
Kepler's SNR (This Work) & 0842550101 & Mar. 2020 & (262.65359, -21.500305) & 92.67 & 132.0 \\
Background (This Work) & 0051940401 & Mar. 2001 & (277.10040, -11.826722) & 92.36 & 9.8 \\
\enddata
\tablecomments{The roll angle is the averaged value of those of both RGS detectors because they are $\sim$0.15~degree different from each other.}\label{tab:obs}
\end{deluxetable*}

Table \ref{tab:rgs_obs} shows a list of RGS observations of Kepler's SNR. 
Because the difference of the roll angle is too large to allow the observations to be merged and the cross-dispersion directions do not correspond to the same portions of Kepler's SNR, we only use the longest observation {\tt 0842550101} in this paper. 
This observation shows almost no background flares in its light curve, so the full exposure is available for science. 
We also use a blank-sky observation {\tt 0051940401} as the detector background for our analysis due to the brightness and extent of the source. 
All observation data are screened by {\tt cifbuild} and {\tt odfingest} commands in the {\tt SAS 18.0.0} package \citep{Gabriel04}. 
In order to divide the analysis regions and create spectral response files, we use the ACIS-S image in the 470--860~eV (14.4--26.4~\AA) band obtained with the {\it Chandra} satellite \citep{Garmire03} in 2006 (Obs. ID: {\tt 6714}--{\tt 6718} and {\tt 7366}), which we reprocessed using the {\tt chandra\_repro} command and merged using {\tt merge\_obs} from the {\tt CIAO 4.9} package \citep{Fruscione06}.

\section{Spectral Analysis and Results} \label{sec:analysis}

\subsection{Regions and reprocessing}

\begin{figure*}[!ht]
\gridline{\fig{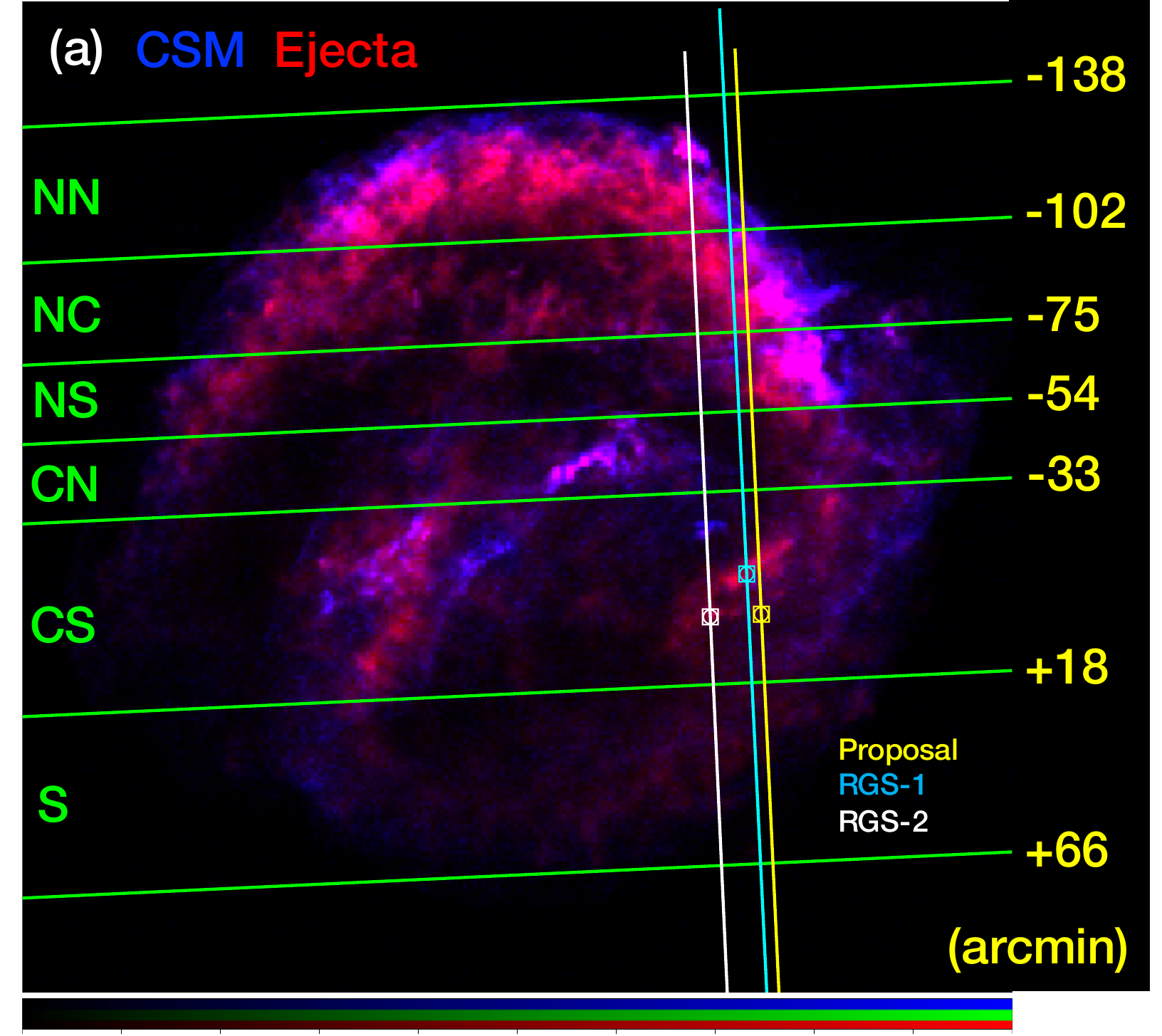}{0.58\textwidth}{}}
\gridline{\fig{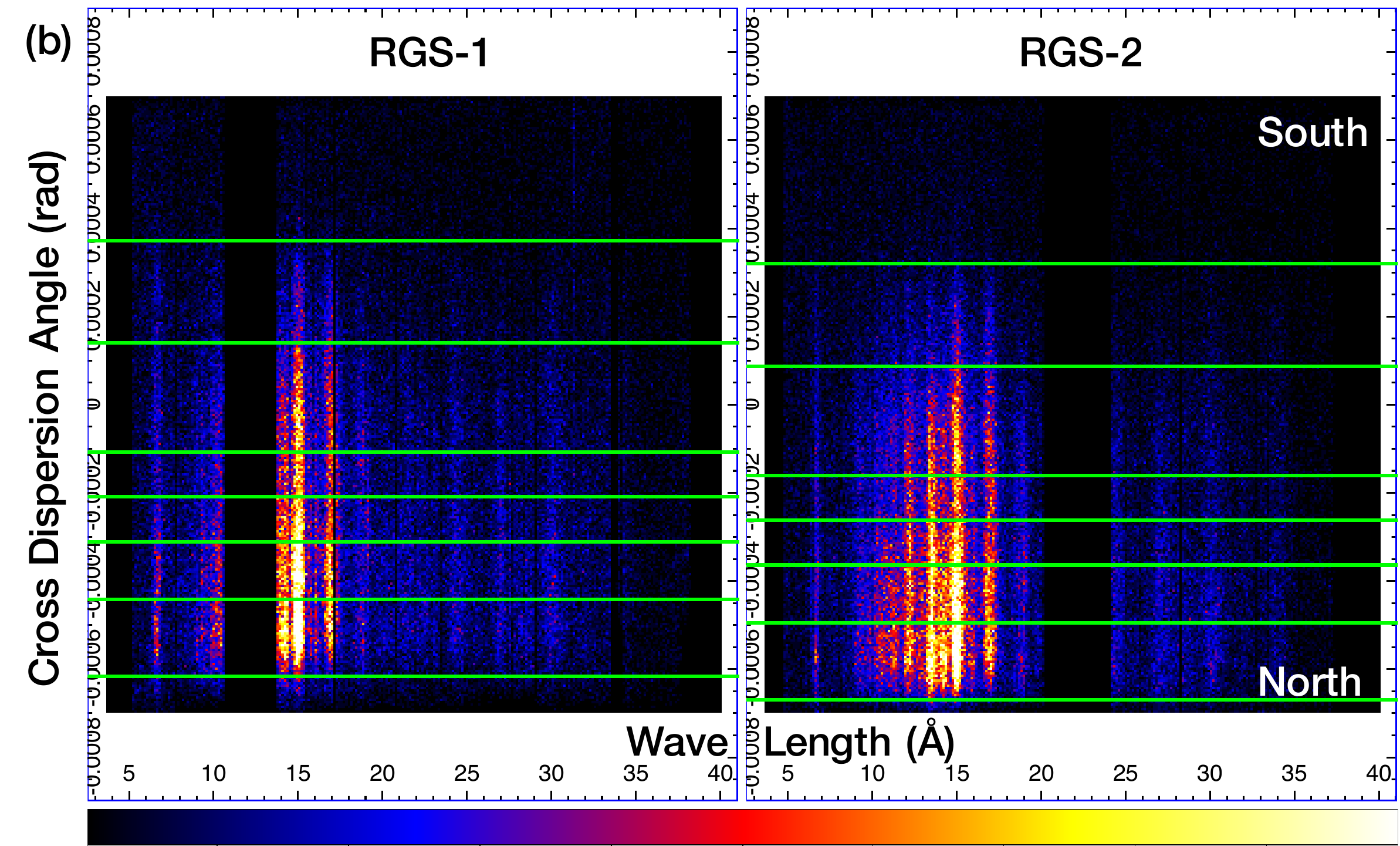}{0.63\textwidth}{}}
\caption{(a) Analysis strips on the {\it Chandra} ACIS-S image of Kepler's SNR in linear scale. The blue and red represent the CSM band (470--600~eV) and the ejecta (Fe-L: 700--860~eV), respectively. Notice that the O Ly$\alpha$ line is not included in both of them. The yellow, cyan, and white color mean distinction of the proposal value and the actual values of both RGS detectors. The squares and vertical lines represent their pointing position and cross dispersion axes, respectively. The green horizontal lines represent dividing positions of each strips and the yellow values are their cross dispersion angle on the proposal axis, whose roll angle is shown in Table \ref{tab:obs}. (b) Strips on the both RGS images. Both color scales are linear in order to emphasize the spacial distribution of the brightness.}\label{fig:reg}
\end{figure*}

For spatially resolved spectroscopy, we divide the field of view along the dispersion angle of the RGS. 
The events detected along the dispersion angle contain a mix of positional information and incident photon energy.
The extraction regions perpendicular to the dispersion axis contain information from several, independent strips covering the SNR.
Figure \ref{fig:reg} (a) shows the 6 strips of our analysis, where we consider various  geometric features of the SNR and the $\sim$5~arcsec angular resolution of the {\it XMM-Newton} telescopes \citep{Jansen01}. 
We identify each strip by name and classify them into 3 groups. 
Strip N*s include the northern bright rim of the remnant. 
The central bar structure lies in strip C*s, but strip CS also has an edge structure in the western half. 
The strip S represents the dark southern structure. 
Note that the effects of vignetting by the  {\it XMM-Newton} telescopes is less than 5\%, given that the diameter of Kepler's SNR is $\sim$4~arcmin. 
Therefore we ignore vignetting in our analysis. \par
To reprocess the {\it XMM-Newton} observation, we use the {\tt rgsproc} command in {\tt SAS}. 
For extracting events in each strip, we consider the slight difference of the pointing center of the two RGS detectors, which cannot be ignored for such a source distribution as complicated as that of Kepler's SNR. 
Figure \ref{fig:reg} (b) shows the strip division on the RGS raw images after accounting for the pointing shift. 
We extract events in the energy band 470--860~eV, which includes the CSM-indicator line emission (N Ly$\alpha$), CSM-dominated lines (O He$\alpha$ \& O Ly$\alpha$), and ejecta-dominated lines (Fe-L lines).  
Note that we exclude the Ne and Mg lines because their energy band could include more complicated ejecta components \citep[see Figure 2 in][]{Katsuda15}. 
We sum up the spectra in the 1st order from each RGS detector. 
We exclude some bad spectral bins with extremely poor photon statistics, due to bad CCD pixels of the RGS detectors.

\subsection{Fitting results} \label{sec:fitting}

\begin{figure*}[!ht]
\gridline{\fig{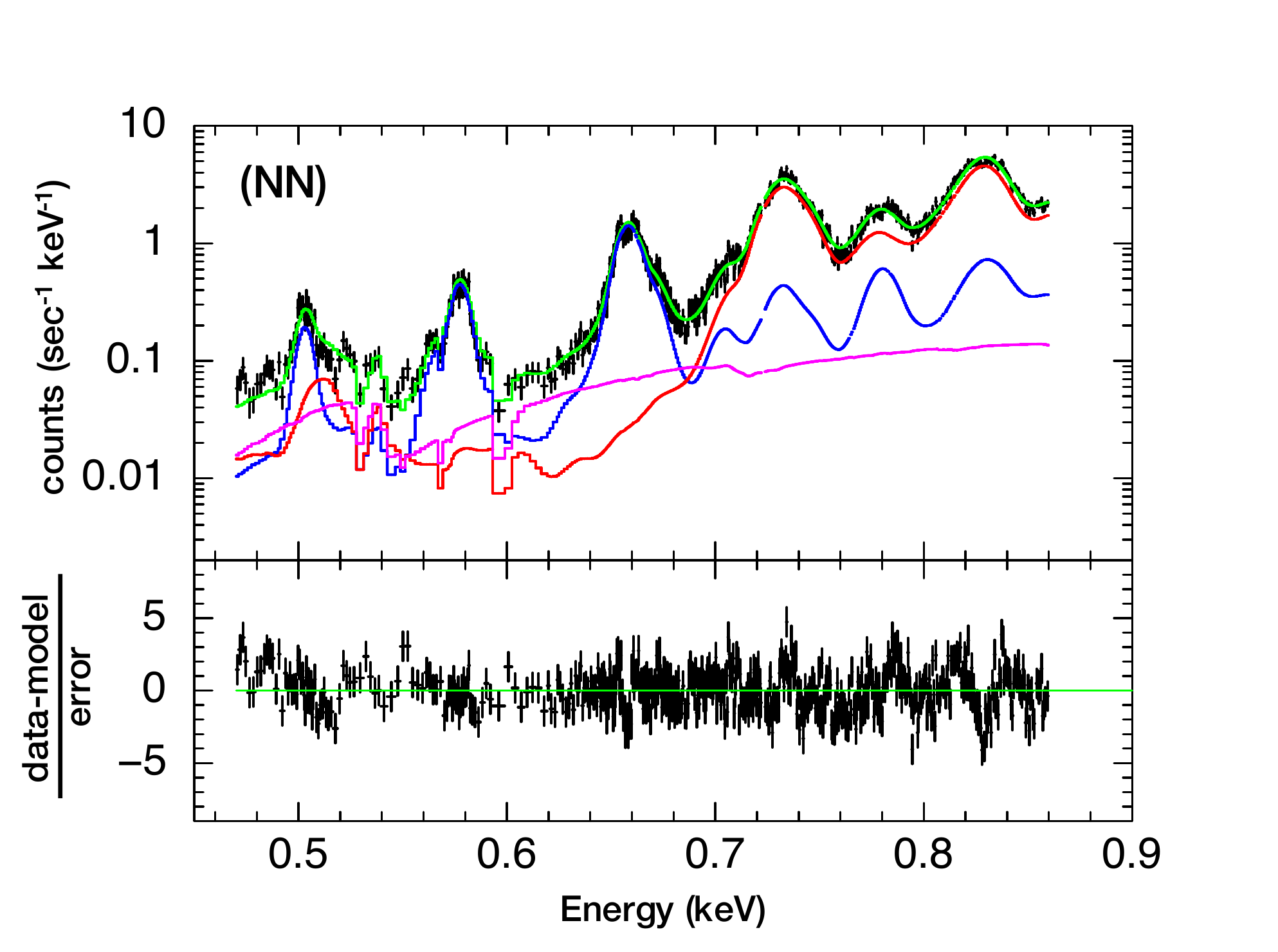}{0.48\textwidth}{} \fig{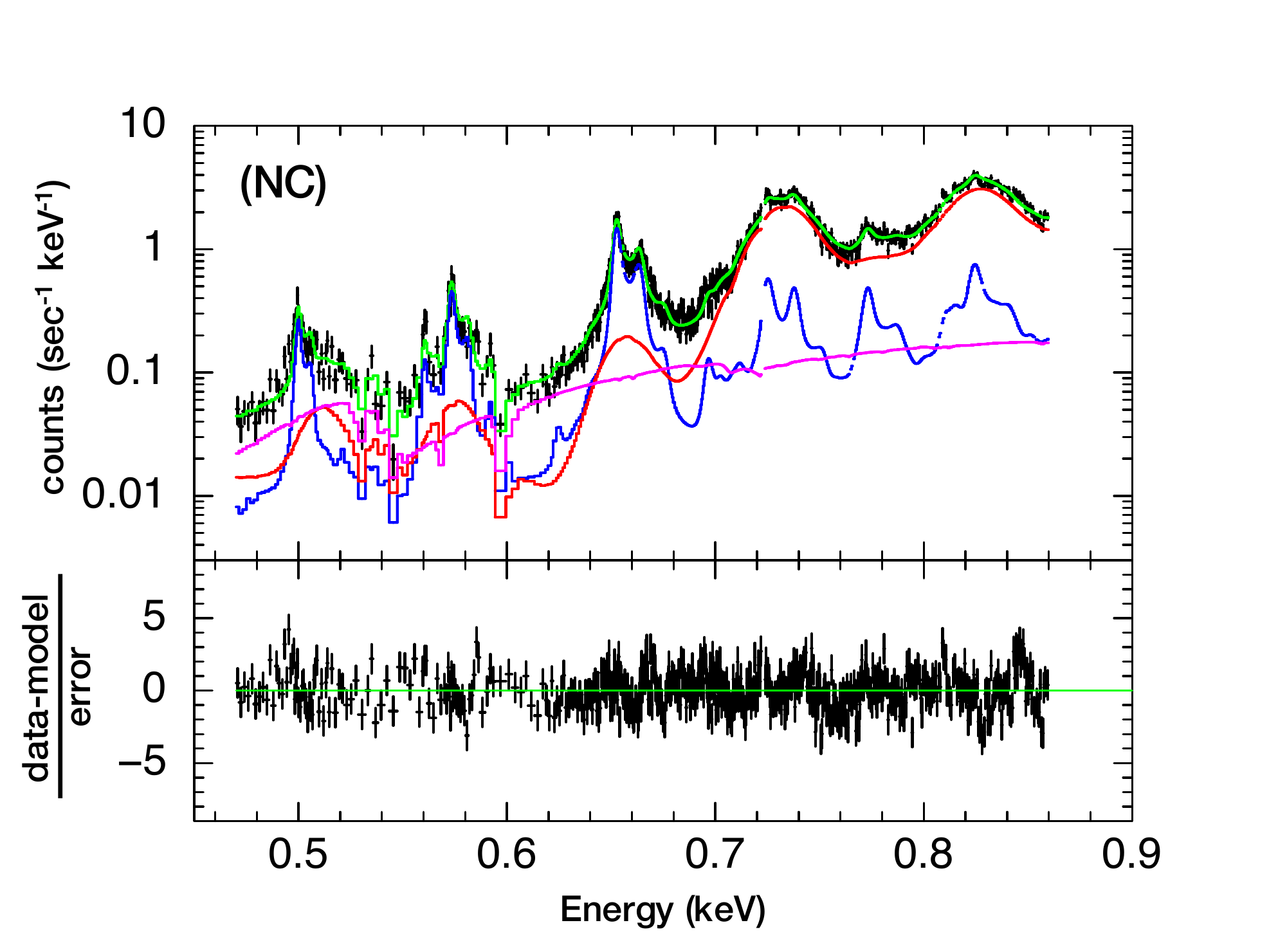}{0.48\textwidth}{}} 
\gridline{\fig{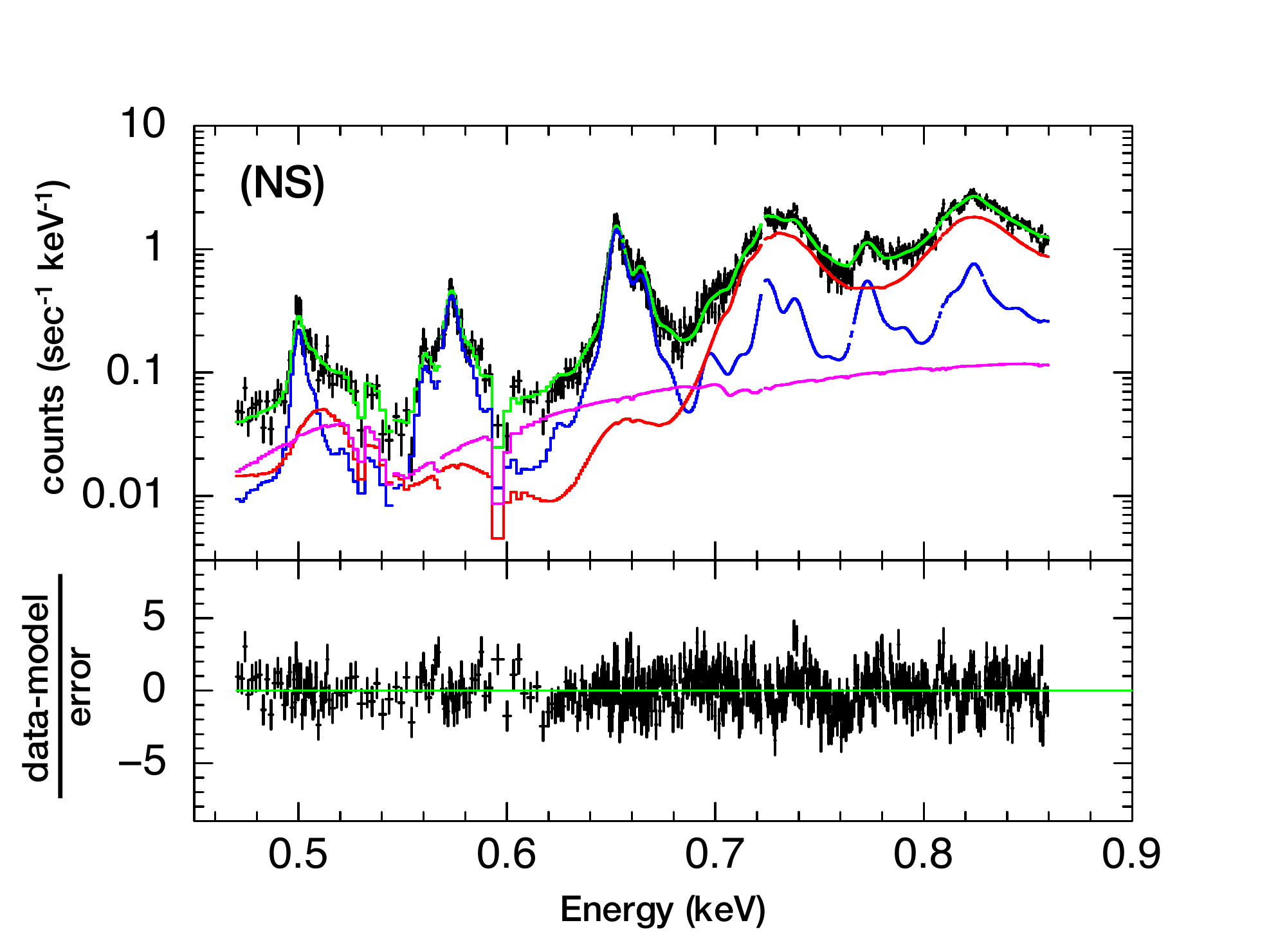}{0.48\textwidth}{} \fig{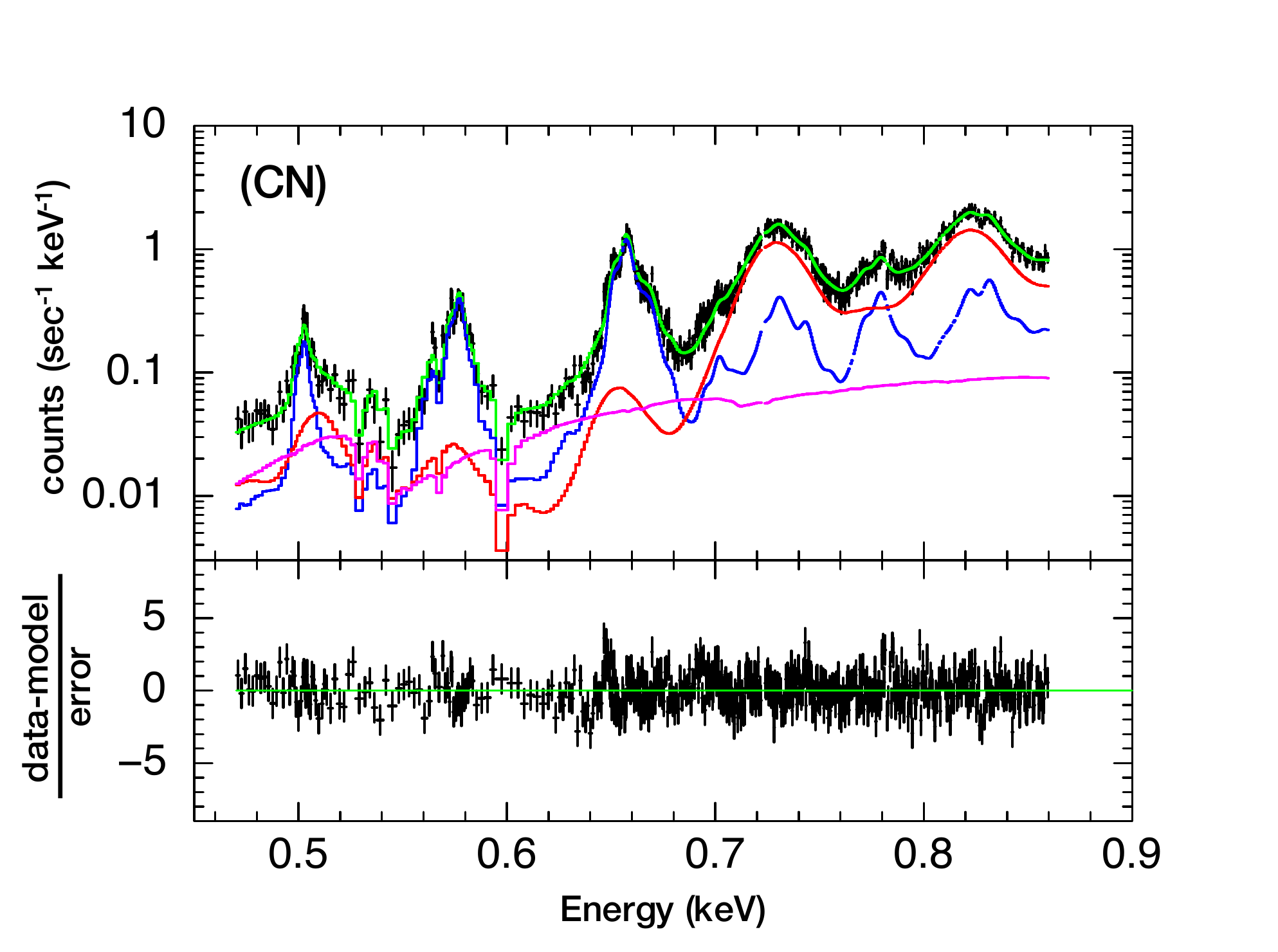}{0.48\textwidth}{}}
\gridline{\fig{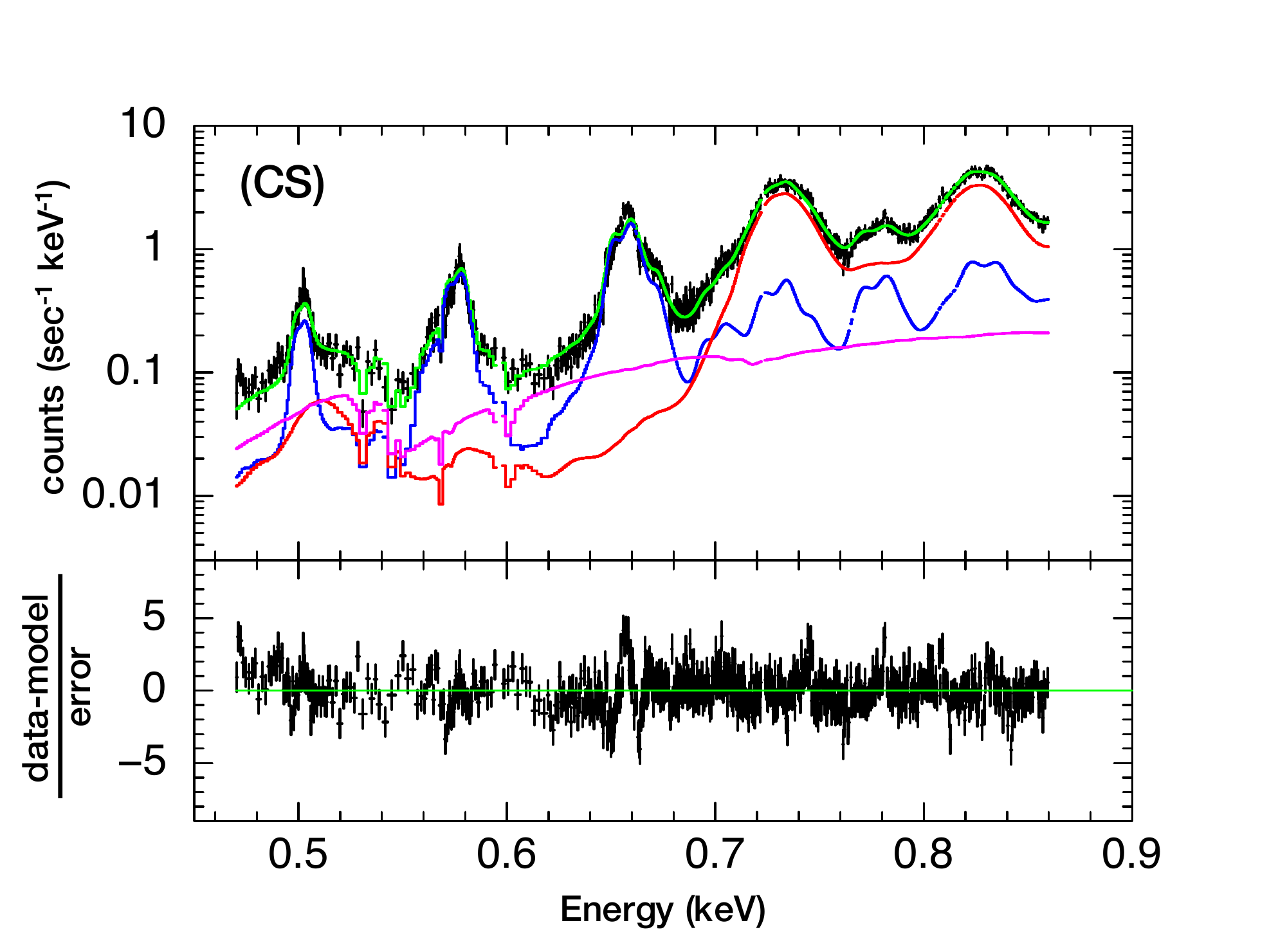}{0.48\textwidth}{} \fig{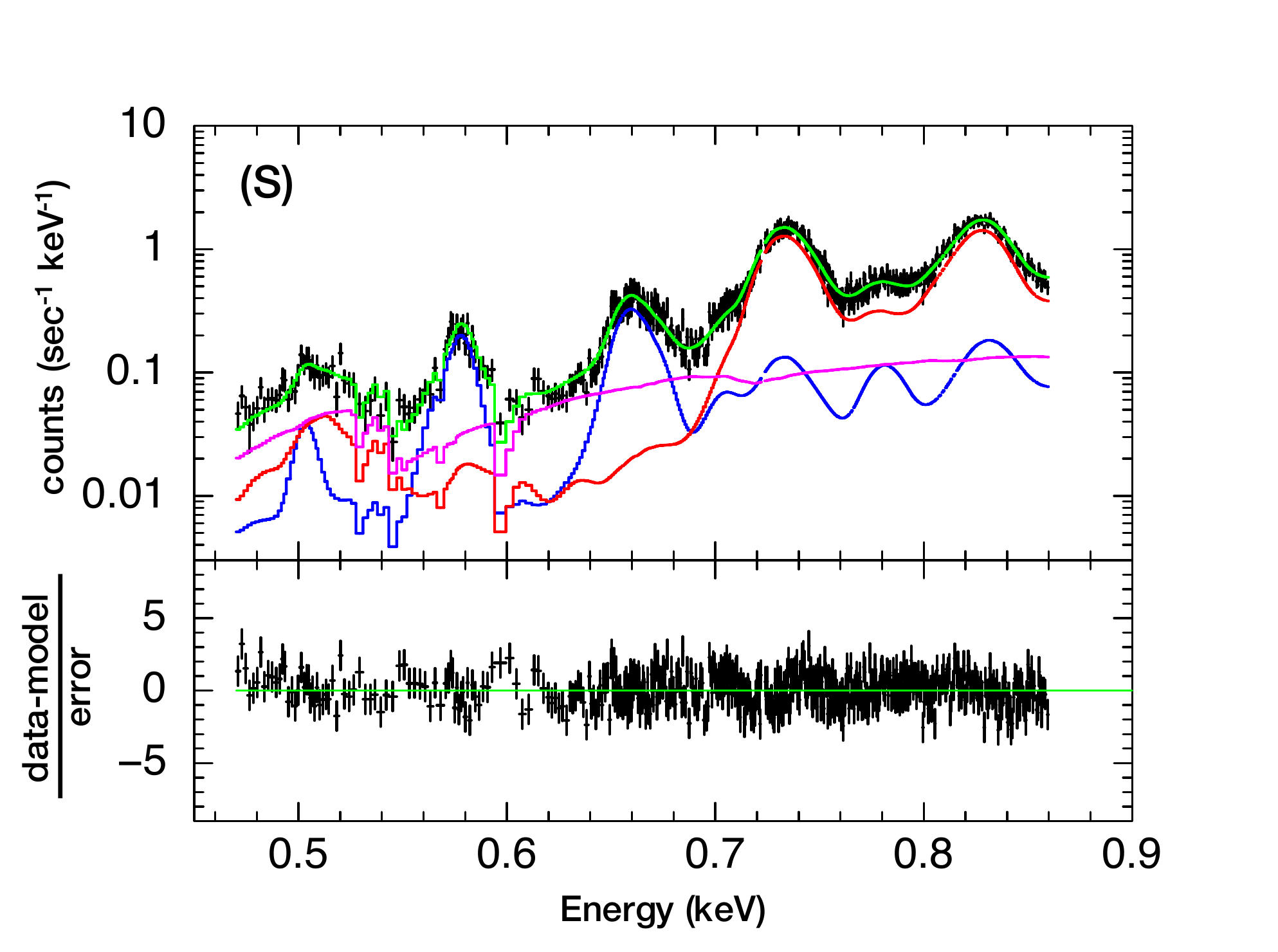}{0.48\textwidth}{}}
\caption{Fitting Results in the northern strips. Blue lines are {\tt vpshock} components representing CSM, and red lines are {\tt vnei} of ejecta. Magenta lines are the {\tt PL} component. Green lines are the sum of them, and lower panels show the residual of our fitting, which we calculate the values of {\tt (data-model)/error} in W-statistics \citep{Kaastra17}.}\label{fig:result}
\end{figure*}

We fit the spectra using the {\tt Xspec 12.9.1} software package \citep{Arnaud96} with {\tt AtomDB 3.0.8} \citep{Foster12}. 
Because un-binned spectra are used, we optimise the emission models using the W-statistic, which is a modification of the C-statistic \citep{Kaastra17} that {\tt Xspec} automatically adopts when including background data.
The fitting model consists of two optically thin thermal emission models, {\tt vpshock} for the CSM component, with the initial value of the ionization timescale $n_{\mathrm{e}} t$ is set to 0, and {\tt vnei} for the ejecta component \citep{Borkowski01}.
Because the detector response created by {\tt rgsrmfgen} command in {\tt SAS} is for a point source, we convolve the response with the spatial emission profile along each dispersion region, by employing the {\tt rgsrmfsmooth} command in the {\tt HEASoft 6.21} package \citep{Blackburn95}. 
As input images, we use {\it Chandra} images for each emission component: 460--600~eV for the CSM and 700--860~eV for the ejecta component (Figure \ref{fig:reg} (a)).
For the CSM component we adopt solar abundances, except for nitrogen, which is allowed to be optimised.
For the ejecta component we use pure metal abundances, following the previous study of \citet{Katsuda15}. \par
Each of these two components are Doppler shifted using the {\tt vashift} model, to account for bulk velocities along the  line-of-sight directions. 
The red-shift parameters of the {\tt vpshock} and {\tt vnei} models are set to be 0, because non-zero redshifts include cosmological effects that affect the model normalisation parameters.
Both components are also each broadened with the {\tt gsmooth} model, to account for both the thermal Doppler broadening and the spread in plasma velocities along the line-of-sight in each strip.
Note that the interpretation of the ionization time scale in the fitting models between {\tt vpshock} and {\tt vnei} is different. 
In the {\tt vpshock} model there is a linear distribution of ionisation ages ($n_{\rm e}t$), whereas the {\tt vnei} model assumes a single ionisation age for the entire plasma \citep{Borkowski01}. \par
Concerning the element abundances, we assume that N emission originates solely from the CSM component, so we allow only the N abundance to be free for the CSM component; the other abundances are set to their solar values \citep{Wilms00}. The N abundance of the ejecta component is fixed to 0.
Because we assume the pure-metal ejecta, we set abundance ratios to H for the intermediate mass elements (Si, S, Ar, and Ca) to be fixed to $10^5$, the iron-group elements (Fe and Ni) to be free near $10^5$, and treated O, Ne, and Mg abundances as free parameters. 
Note that the K-shell emission lines of the intermediate mass elements are outside the RGS spectral band.
We also include a {\tt Power Law} component ({\tt PL}) for the synchrotron radiation, which is thought to be the dominant component of the non-thermal continuum emission in this energy band \citep{Nagayoshi20}. 
The {\tt PL} photon index is defined as $\Gamma$, where the differential number of photons goes as $N(E) \propto E^{-\Gamma}$ as a function of energy $E$. In this paper, $\Gamma$ is fixed to 2.64, the best-fit value in \citet{Katsuda15}, which is also broadly consistent with the results of other studies (2.2--2.6 in \citealt{Bamba05}, 2.60 in \citealt{Park13}, and 2.63$\pm$0.04 in \citealt{Nagayoshi20}).
Finally, the interstellar absorption is modelled with the {\tt tbabs} model \citep{Wilms00}, and is applied to all model components. 
Because there is some discrepancy in published values for the absorption column density $N_{\mathrm{H}}$ toward Kepler's SNR (for example, $N_{\rm H}=5.2 \times 10^{21}$~cm${}^{-2}$ in \citealt{Reynolds07} and $6.4 \times 10^{21}$~cm${}^{-2}$ in \citealt{Katsuda15}), we set $N_{\mathrm{H}}$ to be a free parameter. 
In summary, our fitting model is {\tt tbabs*(vashift*gsmooth*vpshock+vashift*gsmooth*vnei+PL)}. \par
Figure \ref{fig:result} shows our fitted results for each extracted strip. 
The lower panels show the {\tt (data-model)/error} value, i.e. the fitting residuals, which in general show no obvious feature except for strip CS.
These results indicate that our fitting model well-reproduces the observational data, with a relatively simple model in which the CSM and ejecta emission is each represented with a single non-equilibrium ionisation emission model.
Note that, as a check, we also fitted the entire RGS energy band (0.4--2.5~keV). 
The main results in the strips with more compact emission features (NC, NS, and CN) are consistent with the results for the narrower energy band (Table \ref{tab:results}). 
Fits for the other strips are unreliable especially in the narrow energy band we focus here, which may be due to the complicated structure of them for such a 1-D detector.

\subsection{Strip CS}

\begin{figure*}[!ht]
\gridline{\fig{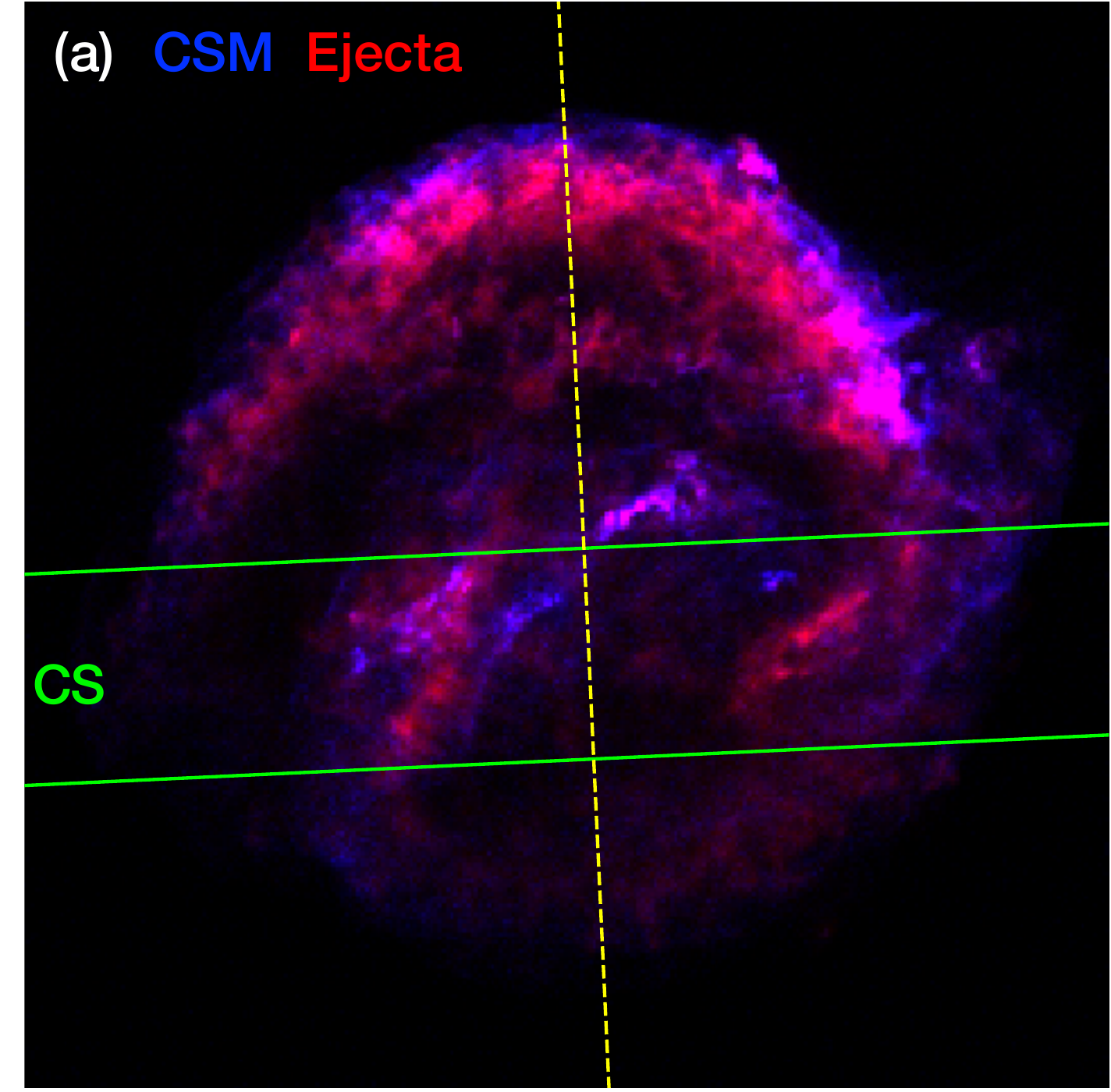}{0.4\textwidth}{} \fig{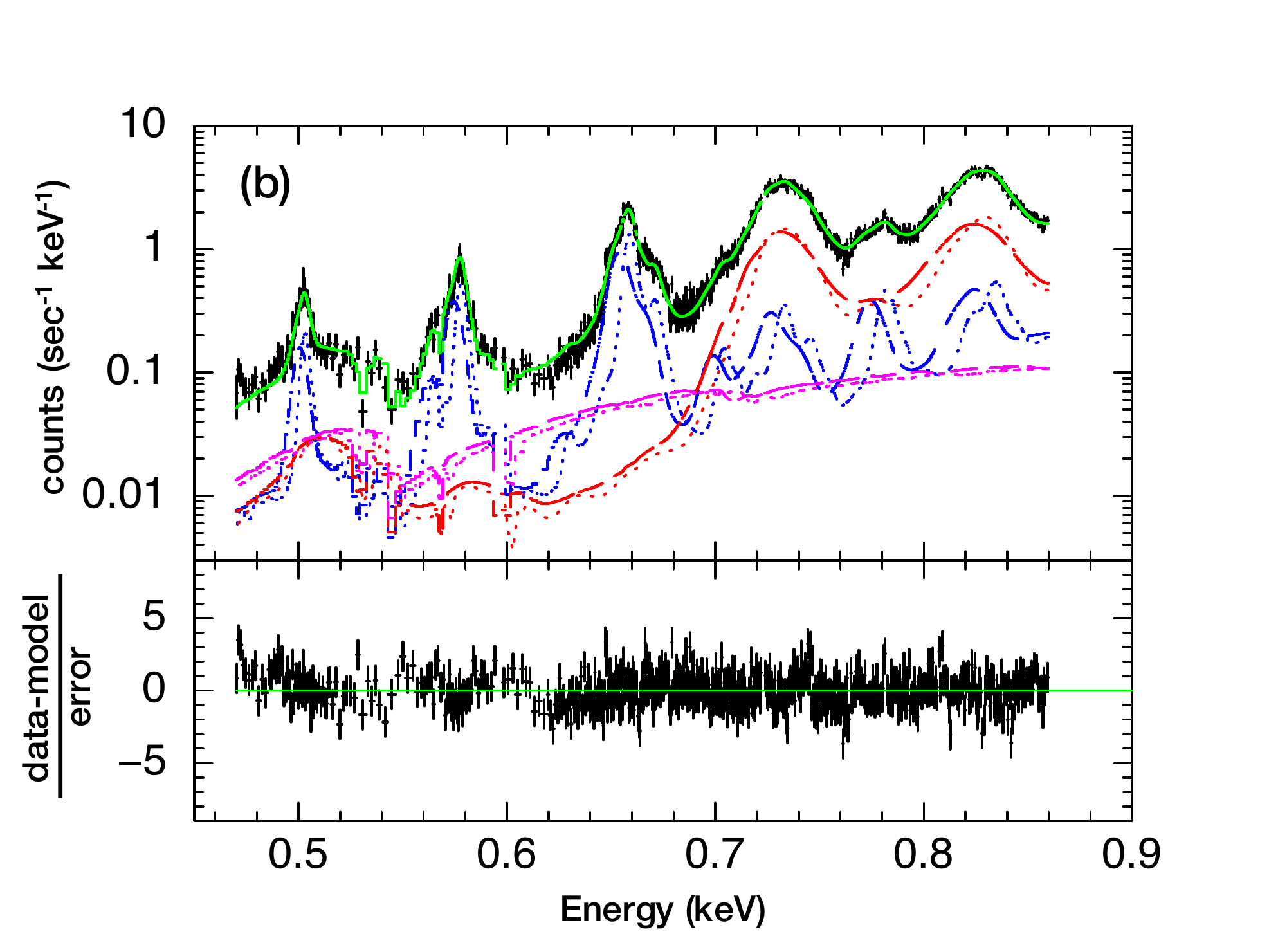}{0.6\textwidth}{}}
\caption{(a) Yellow dashed line represents the division boundary between the west and east halves of the remnant. (b) Fitting results in the strip CS with the improved model. Dashed lines represent the components from west half and dots do east. Other legends are same as Figure \ref{fig:result}.}\label{fig:reg2}
\end{figure*}

\begin{figure*}[!ht]
\gridline{\fig{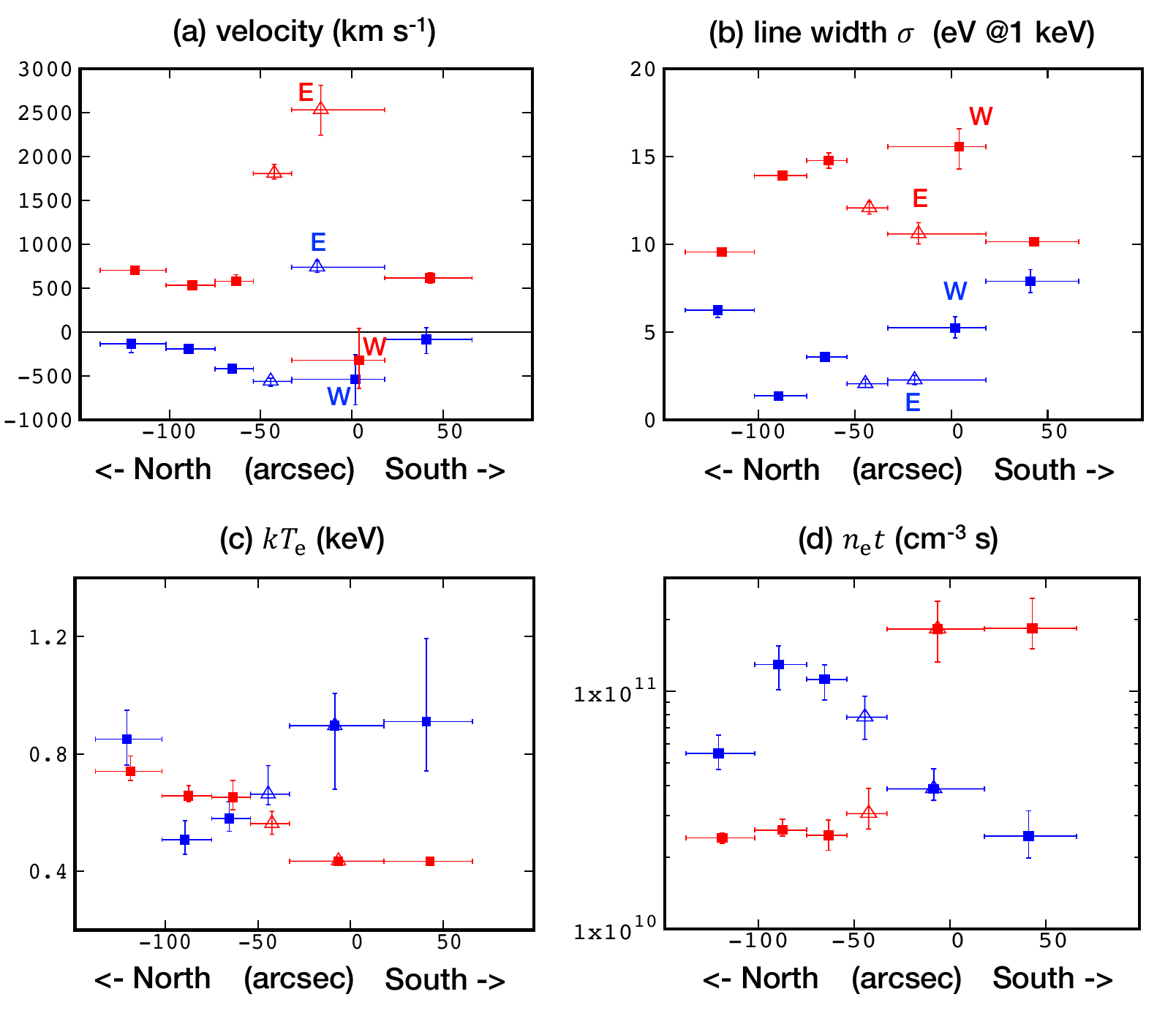}{1\textwidth}{}}
\caption{Results of the free parameters of our fitting with 1$\sigma$ confidencial level. The horizontal axis is the cross-dispersion angle in the unit of arcsec, which is shown in Figure \ref{fig:reg}. (a) Line-of-sight velocity. The positive values mean red-shifted, i.e., moving away from us. The blue points show CSM parameters and red show ejecta ones. The filled square points represent components in the outer rim and outlined triangle points do in the central bar. The black solid line means 0 km s${}^{-1}$. For the strip CS, we distinguish the plots from the west and east halves. (b) Same but line width $\sigma$ converted at 1~keV. (c) Electron temperature $kT$ of {\tt vpshock} and {\tt vnei} models. (d) Same but ionization parameter $n_{\mathrm{e}}t$}.\label{fig:plot1}
\end{figure*}

\begin{deluxetable*}{cccccccc}[!ht]
\tablecaption{Best-fit parameters}
\tabletypesize{\small}
\tablewidth{0pt}
\tablehead{
 & \colhead{NN} & \colhead{NC} & \colhead{NS} & \colhead{CN} & \colhead{CS-E} & \colhead{CS-W} & \colhead{S}\\
Region & Edge & Edge & Edge & Center & Center & Edge & Edge}
\startdata
CSM & & & & & & & \\
\hline
velocity (km s${}^{-1}$) & $-136^{+44}_{-109}$ & $-192^{+29}_{-29}$ & $-419^{+38}_{-41}$ & $-562^{+36}_{-56}$ & $738^{+75}_{-60}$ & $-539^{+280}_{-290}$ & $-86^{+135}_{-160}$ \\
line width $\sigma$ (eV @ 1 keV) & $6.3^{+0.1}_{-0.4}$ & $1.4^{+0.2}_{-0.2}$ & $3.6^{+0.2}_{-0.2}$ & $2.1^{+0.3}_{-0.2}$ & $2.3^{+0.3}_{-0.3}$ & $5.2^{+0.6}_{-0.6}$ & $7.9^{+0.7}_{-0.6}$ \\
$kT_{\mathrm{e}}$ (keV) & $0.85^{+0.10}_{-0.09}$ & $0.51^{+0.07}_{-0.05}$ & $0.58^{+0.06}_{-0.04}$ & $0.66^{+0.10}_{-0.04}$ & \multicolumn{2}{c}{$0.90^{+0.11}_{-0.22}$} & $0.91^{+0.28}_{-0.17}$ \\
$n_{\mathrm{e}}t$ (10 ${}^{10}$~cm${}^{-3}$ s) & $5.5^{+1.0}_{-0.8}$ & $12.9^{+2.5}_{-2.8}$ & $11.2^{+1.7}_{-2.0}$ & $7.8^{+1.7}_{-1.5}$ & \multicolumn{2}{c}{$3.9^{+0.8}_{-0.4}$} & $2.5^{+0.7}_{-0.5}$ \\
N (solar) & $1.51^{+0.18}_{-0.11}$ & $2.04^{+0.19}_{-0.19}$ & $1.77^{+0.16}_{-0.16}$ & $1.41^{+0.14}_{-0.14}$ & \multicolumn{2}{c}{$1.54^{+0.11}_{-0.10}$} & $0.77^{+0.23}_{-0.23}$ \\
other elements (solar) & 1 & 1 & 1 & 1 & \multicolumn{2}{c}{1} & 1 \\
\hline
Ejecta & & & & & & & \\
\hline
velocity (km s${}^{-1}$) & $702^{+33}_{-34}$ & $532^{+53}_{-47}$ & $578^{+75}_{-38}$ & $1805^{+104}_{-63}$ & $2531^{+280}_{-289}$ & $-321^{+361}_{-323}$ & $615^{+61}_{-59}$ \\
line width $\sigma$ (eV @ 1 keV) & $9.6^{+0.1}_{-0.1}$ & $13.9^{+0.3}_{-0.2}$ & $14.8^{+0.4}_{-0.4}$ & $12.1^{+0.4}_{-0.4}$ & $10.6^{+0.7}_{-0.6}$ & $15.6^{+1.0}_{-1.3}$ & $10.1^{+0.3}_{-0.2}$ \\
$kT_{\mathrm{e}}$ (keV) & $0.74^{+0.05}_{-0.03}$ & $0.66^{+0.04}_{-0.02}$ & $0.65^{+0.06}_{-0.04}$ & $0.56^{+0.04}_{-0.04}$ & \multicolumn{2}{c}{$0.43^{+0.01}_{-0.00}$} & $0.43^{+0.01}_{-0.00}$ \\
$n_{\mathrm{e}}t$ (10 ${}^{10}$~cm${}^{-3}$ s) & $2.4^{+0.1}_{-0.1}$ & $2.6^{+0.3}_{-0.2}$ & $2.5^{+0.4}_{-0.3}$ & $3.1^{+0.8}_{-0.4}$ & \multicolumn{2}{c}{$18.3^{+5.6}_{-5.0}$} & $18.4^{+6.3}_{-3.3}$ \\
H,He,C (solar) & 1 & 1 & 1 & 1 & \multicolumn{2}{c}{1} & 1 \\
N (solar) & 0 & 0 & 0 & 0 & \multicolumn{2}{c}{0} & 0 \\
O,Ne,Mg ($10^{3}$ solar) & $< 1.8$ & $3.9^{+0.7}_{-0.4}$ & $< 1.3$ & $1.5^{+0.5}_{-0.6}$ & \multicolumn{2}{c}{$< 0.3$} & $< 0.3$ \\
Si,S,Ar,Ca ($10^{3}$ solar) & 100 & 100 & 100 & 100 & \multicolumn{2}{c}{100} & 100 \\
Fe,Ni ($10^{3}$ solar) & $117^{+42}_{-8}$ & $128^{+25}_{-25}$ & $77^{+21}_{-15}$ & $66^{+17}_{-11}$ & \multicolumn{2}{c}{$68^{+19}_{-21}$} & $44^{13}_{-5}$ \\
\hline
Total & & & & & & & \\
\hline
$n_\mathrm{H}$ ($10^{21}$~cm${}^{-2}$) & $5.7^{+0.2}_{-0.2}$ & $5.6^{+0.2}_{-0.2}$ & $5.5^{+0.2}_{-0.3}$ & $5.4^{+0.2}_{-0.3}$ & \multicolumn{2}{c}{$5.8^{+0.2}_{-0.1}$} & $5.3^{+0.1}_{-0.2}$ \\
W-stat & 1681.47 & 1602.46 & 1584.98 & 1410.52 & \multicolumn{2}{c}{1323.61} & 1233.72 \\
d.o.f. & 1131 & 1131 & 1131 & 1131 & \multicolumn{2}{c}{1127} & 1131 \\
W-stat/d.o.f & 1.487 & 1.417 & 1.401 & 1.247 & \multicolumn{2}{c}{1.174} & 1.091
\enddata
\tablecomments{Notice that the definition of the ionaization timescale $n_{\mathrm{e}}t$ is different in both {\it vpshock} and {\it vnei} models. Almost all parameters except for velocity and line width in the strip CS are linked between the east- and west-halves.}\label{tab:results}
\end{deluxetable*}

As shown in Figure \ref{fig:result}, we found that the line shape of the O Ly$\alpha$ in CS is not well reproduced by the model explained above. 
This strip obviously includes both the central bar-like structure and the western side of the main shell (Figure \ref{fig:reg}). 
We therefore applied another model particularly for the CS strip, where the RGS spectrum is fitted with a model assuming the superposition of two spatially-distinct components.
Specifically we use the combination of {\tt vpshock} for CSM at the center, {\tt vpshock} for CSM at the rim, {\tt vnei} for ejecta at the center, and {\tt vnei} for ejecta at the rim. 
The Doppler parameters (the line-of-sight velocity and the line width) are set to be independent parameters, but the other parameters are forced to be the same.
We set the normalization parameters to be coupled between the western and the central parts.
This procedure also requires that the response matrix take into account the spatial differences corresponding to the central bar and western part of the shell.
To accomplish this, we split the image of Kepler's SNR (Figure \ref{fig:reg2} (a)) into two, one for the western and one for the eastern half. 
We use the western-half image for the spatial-part of the detector response of the outer rim component and the eastern-half image for the central bar detector response (cf., \citet{Katsuda13}). 
Figure \ref{fig:reg2} (b) shows the model fitting results for this more elaborate, two-spatial-component model.
Compared to the previous result (Figure \ref{fig:result}), the residuals near the peak of the O Ly$\alpha$ emission are considerably reduced. 
The fits are much better with the {\tt W-stat (d.o.f)} values improved from {\tt 1641.43 (1131)} to {\tt 1323.61 (1127)}, and with the F-test probability is much lower than 0.01 based on the chi-square values.
The best-fit parameter table (Table \ref{tab:results}) and plots (Figure \ref{fig:plot1}) include these additional analyses.

\clearpage

\begin{figure*}[!ht]
\gridline{\fig{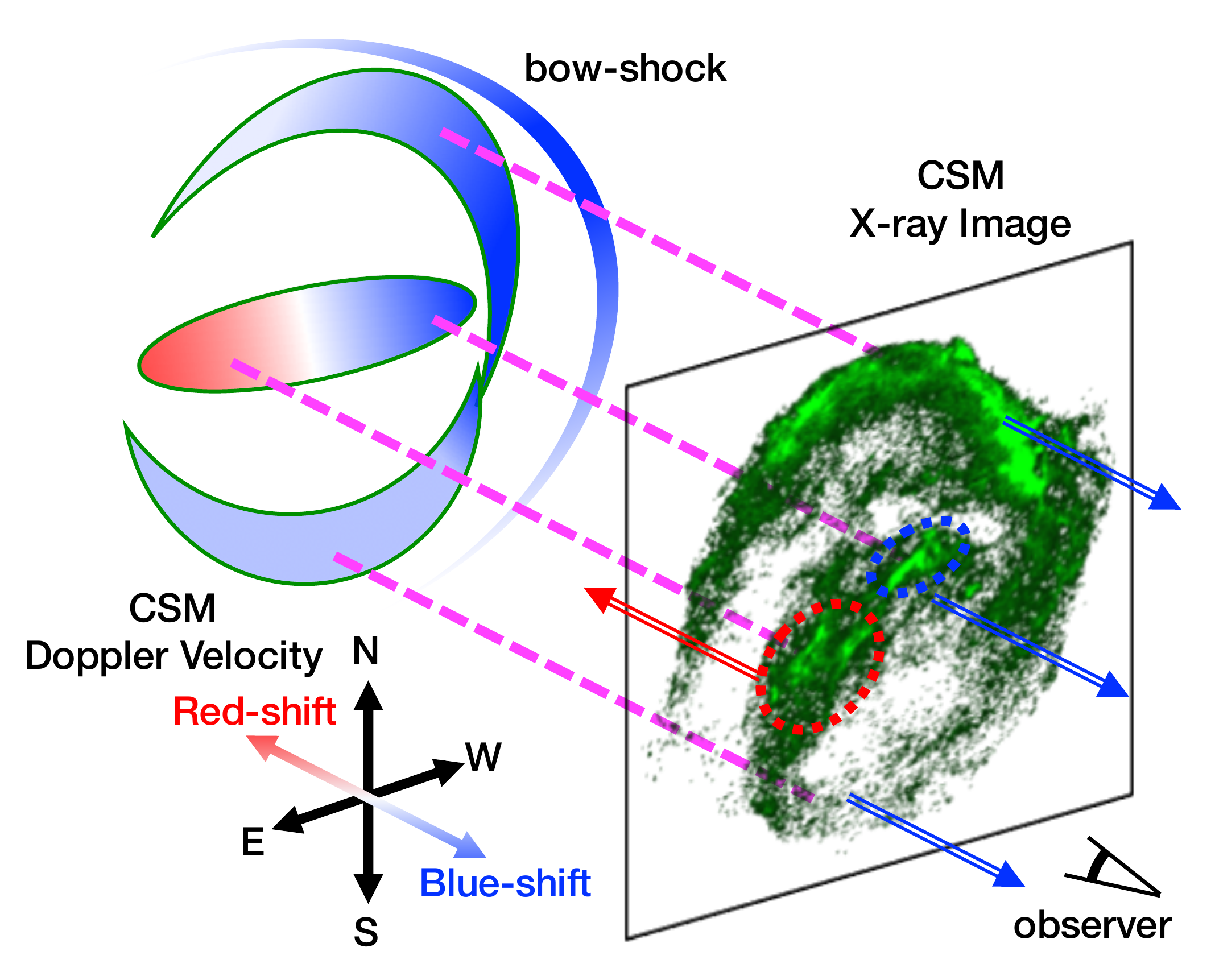}{0.6\textwidth}{}}
\caption{A schematic view of the kinematics of CSM in Kepler's SNR from diagonally above. The blue or red color surrounded by green line represents the Doppler shift of CSM. Darker color means a larger shifted component. We also add a bow-shock generated by the ``runaway'' AGB star. The green image is the X-ray image of CSM (same as Figure \ref{fig:reg}) and colored arrows also mean the Doppler shift at each position. }\label{fig:schematics}
\end{figure*}

\section{Discussion} \label{sec:discussion}

The modelling results give us insight in both the velocity structure and composition of the CSM and the ejecta component.
Figures \ref{fig:plot1} (a) and (b) show the Doppler velocities and widths for both these components.
For the velocity structure it is important to remember that for a perfectly spherical expanding shell, no net Doppler shift is expected. 
So any Doppler shift is either due to an overall Doppler shift for the system, which for Kepler is expected to be 230$\pm$50~km s${}^{-1}$ \citep{Minkowski59}, or due to deviations from spherical symmetry. 
At the edges of the expanding shell, all motions are in the plane of the sky, and deviations from spherical symmetry should not play a role --- except for cases where the emission comes from a wider region, or if there is some contamination due the instrumental point spread function. \par
Doppler widths can have several causes: i) the bulk velocity of the expanding shell, ii) motions from unresolved knots or turbulence, and iii) thermal Doppler shifts.
For a spherically expanding shell, the bulk velocity broadening is largest in the center where the line-of-sight motions have maximum approaching and receding components toward the observer. 
At the edges of the SNR the bulk motion is in the plane of the sky, and hence bulk velocity broadening should be minimal. 
However, turbulent motions and thermal Doppler broadening should to first order not be dependent on the projected location on the sky. \par
In particular, thermal Doppler broadening received some attention, as collisionless shocks likely heat the different plasma components to different temperatures, instead of equipartition of temperatures \citep[see][and references therein]{Ghavamian13,Vink15}. 
In the most extreme case the temperatures are given by
\begin{equation}
kT_i= \frac{3}{16}m_i V_{\rm S}^2,
\end{equation}
with $T_i$ the temperature of a given species $i$, $m_i$ the species' particle mass, and $V_{\rm S}$ the shock velocity.
Post-shock Coulomb interactions, may then equilibrate the temperatures further downstream of the shock, proportional to $n_{\mathrm{e}}t$. \par
Below we will discuss separately the velocity structure and composition of the CSM and ejecta components.

\subsection{The circumstellar medium component}

\subsubsection{Doppler velocity structure}

Focusing on CSM velocity parameters at the edge of the remnant, the line-of-sight velocity is fit in the narrow range between 0--500~km s${}^{-1}$ toward us. 
These results can be generally explained by the bulk motion of the whole remnant. From an observation of some bright optical knots, the radial velocity of the remnant is estimated to be blue-shifted with a velocity of 230$\pm$50~km s${}^{-1}$ \citep{Minkowski59}.
Combining the observation of the tangental velocity, \citet{Bandiera91} estimated the 3-Dimensional kinematics of the remnant as it moves towards the northwest and in our direction along the line-of-sight with a space velocity of 278$\pm$12~km s${}^{-1}$. 
Previous studies attributed the overall blue-shifts and proper motions to be caused by the stellar wind from a ``runaway" star with a large system motion \citep{Bandiera87,Borkowski92}. Original binary system could be like the Mira’s binary, whose velocity is $\sim$100~km s${}^{-1}$ and who has a bow-shock in the infrared and ultraviolet band \citep[cf.,][]{Martin07}. Such ``runaway'' massive stars are still found even 600~pc above the galactic plane like Kepler's position \citep[e.g.,][]{Martin06}. \par
In the central region, we can see a red-shifted feature in the north-west structure (strip CN) and a blue-shifted structure in the south-east (the east half of strip CS). 
These results are consistent with the previous optical observation of the H$\alpha$ lines from CSM knots in this vicinity \citep{Blair91}. 
\citet{Blair91} compared the centroid energy of the broad H$\alpha$ lines and estimated the difference as $\sim$20~\AA~between these CSM knots in the area we are discussing \citep[see Table 3 in][]{Blair91}. 
The line widths in these regions are narrower than that in the edge (Figure \ref{fig:plot1} (b)). 
These results are explainable by a dense and torus-like CSM, which \citet{Burkey13} indicated using a hydrodynamical simulation. 
Such a torus-like CSM is explained by the SD scenario of type Ia SN \citep{Hachisu08}, and is also found in other SNe and SNRs like SN2012dn \citep{Nagao17,Nagao18} and N103B \citep{Yamaguchi21}, for example. \par
Figure \ref{fig:schematics} shows our interpretation of the velocity structure of the shocked CSM. 
This figure shows that the bipolar CSM to the north and south with a central dense torus \citep{Burkey13} is moving towards the northwest and in our direction. 
The angle to us out of the sky plane is estimated as $\sim$35~degree \citep{Bandiera91}. 
Such movement is because of the ``runaway'' AGB star, which generates a bow-shock in the direction of motion \citep{Borkowski92}. 
The blue-shift of the CSM is expected to follow this motion of the bow-shock. 
Note that the radial velocity is anti-correlated with the density of ISM, for example the northern, bright region moves more slowly \citep{Katsuda08}. 
The central torus is along the axis of the motion of the AGB star. \par
\citet{Borkowski92,Velazquez06,Chiotellis12} did 2D-simulations considering a ``runaway'' AGB star with an isotropic CSM, and \citet{Burkey13} assumed an anisotropic CSM in a static system. 
Combining these ideas, \citet{Toledo-Roy14} carried out 3D-simulations with both assumptions and succeeded in generating the northern bright structure and the central bar of this remnant, that was visible in its X-ray emission (see Figure 10 in \citet{Toledo-Roy14}). 
Future simulations concentrating on the line-of-sight velocity of the CSM could test the validity of this assumption. 
On the other hand, especially based on the CD scenario, \citet{Chiotellis20} generated Kepler's characteristic ``ears'' \citep{Tsebrenko13} with the northern brightness by the PNe-like CSM formed just before the explosion of a WD with a ``runaway'' AGB star. 
Our observed velocity structure in the central region suggest the ears may be a torus-like structure. \par
Remarkably, the line width of the CSM component peaks in the outer two regions, where we expect the broadening effects of the shell velocities to be minimal.
For the northern edge we measured $\sigma=3.6$~eV at 1 keV.
If we interpret the broadening as being purely due to unequilibration of temperatures, the implied shock velocity is $V_{\rm s}=\sqrt{16/3}\sigma_v\approx 2500$~km~s$^{-1}$, which is consistent with the slower proper motions in that region. 
For the southern part  we have $7.9\pm0.7$~eV at 1 keV, corresponding to $V_{\rm s}\approx 5400$~km\ s$^{-1}$, which is somewhat higher than inferred from the measured proper motion for the southern part of the remnant, $4500d_5$~km~s$^{-1}$ \citep{Vink08,Katsuda08}.
Note that the line width provides an upper limit to the shock velocity. 
The relation between ion temperature and velocity dispersion is $kT_i=m_i \sigma_v^2$.
So our measurement for the southern part implies for oxygen, $kT_{\rm O} \sim 900$~keV. \par
However, it is remarkable that for the more central strips --- where we expect  bulk motions to contribute more --- that the line broadening is much lower, which either requires slower shocks, whether interpreted as bulk motions or thermal broadening, and/or stronger equilibration of ion-electron temperatures.
This region is dominated by the bar-like feature in the SNR.

\subsubsection{Nitrogen abundance}

We confirm the existence of Nitrogen from the CSM components in all strips (Table \ref{tab:results}). 
Our result of the N abundance of CSM is generally 1--2 solar, where \citet{Katsuda15} indicated $2.91^{+0.36}_{-0.32}$ solar in the whole of the remnant and $6.08^{+3.17}_{-2.75}$ solar in the southern region \citep[see Table 2 in][]{Katsuda15}. 
They analyzed the wide-band spectra combined with the old RGS data (Obs. ID {\tt 0084100101} in Table \ref{tab:obs}) and CCD data above 2~keV using the X-Ray Imaging Spectrometer on board {\it Suzaku} \citep{Koyama07}. 
\citet{Nagayoshi20} re-analyzed the same datasets by using same fitting models as \citet{Katsuda15} and their result of N abundance was also smaller than previous one, $1.93^{+0.17}_{-0.15}$ solar \citep[see the appendix and Table 4 in][]{Nagayoshi20}. 
They pointed out the reason was the difference of the version of {\tt AtomDB} used. \citet{Nagayoshi20} used the latest version {\tt AtomDB 3.0.9}, whereas \citet{Katsuda15} used the old version {\tt AtomDB 2.0.2}. 
Our work uses {\tt AtomDB 3.0.8} and the new, deep RGS data and supports the result from \citet{Nagayoshi20}. \par
The nitrogen abundance of 1--2 solar translates into an N/O mass ratio of $\approx$ 0.15--0.3 for  the solar abundance values of \citet{Wilms00} that we use here. 
Associating this ratio with the surface abundance ratio in AGB stars as calculated by \citet{Karakas16} (their Figure 2), the required mass of an AGB star is less than 4~$M_{\odot}$ because the N/O rapidly gets larger than 1 for heavier stars. 
On the other hand, the presence of silicate dust requires a C/O ratio smaller than 1 \citep[see the discussion in][]{Vink17}, which is because a carbon excess will result in CO molecule formation instead of silicate dust formation, and this requires an AGB star with $M>4$~$M_{\odot}$ in the models of \citet{Karakas16}.
We have here an apparent contradiction. 
However, note that the N/O and C/O ratios differ considerably between the initial and final thermal pulses. 
So the question is whether most of the material encountered by the SNR blastwave so far corresponds to the first or the final thermal pulses. 
Another concern is that the models of \citet{Karakas16} are for the evolution of  single stars, whereas the progenitor system of SN1604 must have been a binary system. 
In particular, the CD scenario prescribes a common-envelope phase in the progenitor evolution, which may result in different relations for the N/O and C/O ratios as a function of the initial mass of the AGB star.
Notice that in the strip S, the N abundance is lower than 1 solar. 
Considering the small equivalent width of the N line (Figure \ref{fig:result}) and the wide line width (Figure \ref{fig:plot1} (b)), we may observe the mixture of CSM and ISM there.

\subsection{The ejecta component}

\subsubsection{The Doppler velocity structure} 

Table \ref{tab:results} also includes the results for the Doppler shifts and Doppler widths for the ejecta component. 
For the RGS this component is mainly associated with the Fe-L line emission.
In contrast to the CSM, the ejecta velocity parameters imply that the ejecta are receding from the observer. 
This is in qualitative agreement with  the  analysis of radial velocities of bright Si ejecta knots in Kepler's SNR from \citet{Sato17b} and \citet{Millard20}, who found several knots moving with high velocity, approaching 10,000 km s$^{-1}$.
However, the velocities reported here are much less than this. 
This may reflect the different nature of the features under study. 
The compact knots seen by {\it Chandra} are isolated and their lack of deceleration may be due to their motion through low density gaps in the CSM (see discussion in \citealt{Sato17b}). 
The results reported here pertain to large sections of the SNR, which include a mix of ejecta velocities, both from the front and backside of the shell and likely include material encountering more typical, higher density  regions of the CSM. 
Apart from the NS and CS strip, the Doppler velocities are below 1000~km~s$^{-1}$. 
The highest velocity measured is $\approx 2500$~km~s$^{-1}$ for the east component of the CS strip, which corresponds to the central bar.
Given that the shell is approximately moving at 75\% of the shock velocity expected for a high Mach number shock, we expect shell velocities of roughly 1500--3500~km~s$^{-1}$ based on the overall proper motion measurements \citep{Katsuda08,Vink08}, in agreement with the highest Doppler velocity report here. 
The redshift means that the central bar is on average moving away from us, implying that it is located on the far side of the SNR. \par
For the central bar, \citet{Kasuga18} found diffuse Fe-K structures whose centroid energy is lower than that of the neutral Fe line at 6.4~keV, especially in the east half of the strip CS. 
The energy shifts imply a velocity of $\sim$2,000~km s${}^{-1}$, consistent with this work. 
In contrast, the RGS result shows that the Fe-L ejecta in the western half of the strip CS is the only blue-shifted feature. 
This is in agreement with previous studies concerning the Si ejecta knots \citep{Sato17b} and Fe-K bulk structure \citep{Kasuga18}, which found higher centroid energies in this location.
In fact, \citet{Sato20} found a large structure of Fe ejecta that moves towards us \citep[see Figure 12 in][]{Sato20}. 
It may be surprising to find an overall redshift for the ejecta component, whereas the CSM component appears to be blueshifted. 
One possible explanation is that a higher density of the CSM on the near-side of the SNR may trigger an earlier response of the reverse shock, and result in general in slower expansion of the ejecta, whereas at the farside the ejecta is moving faster, resulting in a net redshift. 
The contrast in densities on the near- and far-side could be the result of the overall angle of the bow-shock shaped structure created by the stellar wind from the progenitor system, which is tilted toward us, given the reported overall blueshifts of the CSM. 
But an additional geometrical axis to be considered is that of a possible bipolar wind from the progenitor system, whose axis may not be aligned with the overall direction of the progenitor motion. \par
There may be additional factors at play. For example, \citet{Reynoso99} found extended HI emission---which is thought to be an indicator of ISM---at the back side of this SNR.  
We should, however, be cautious as the nature of the RGS only allows us to resolve only one spatial direction (north to south), which  leaves us with uncertainty about the other direction (east to west). 
In the future, X-ray calorimeters \citep{Kelly16} enable us to observe with 2-Dimensional angular resolution using the Resolve detector \citep{Ishisaki18} on board the {\it XRISM} \citep{Tashiro18}, the X-IFU \citep{Barret18} on the {\it Athena+} \citep{Nandra13}, and the LXM on the {\it Lynx} satellite \citep{Bandler19}. 
Such detectors will help us to analyze the line-of-sight kinematics of not only the CSM components but also ejecta in detail without the current limitation on spatially-resolved spectroscopy. \par
The line broadening of the ejecta component is substantially larger than for the CSM component: $\approx$ 9.6--15.6~eV at 1 keV, corresponding to $\sigma_v\approx 2850$--$4650$~km~s$^{-1}$. 
Here the expectation is satisfied that the line broadening should be largest in the central regions where the line of sight speeds are maximised for an expanding spherical shell. 
So very likely, the width provides constraints on the expansion velocity of the ejecta shell. 
The proper motions indicate shock velocities of 2000--4500~km~s$^{-1}$, and the maximum plasma velocities are expected to have a
similar range \citep[c.f., models in][]{Chevalier82}. 
The values of $\sigma_v$ are also larger than the velocity shifts, indicating that indeed we have a more or less complete shell, but may have emission differences between the back and front sides of the shell.
There is also very likely additional thermal line broadening. 
The line width at the outer strips provides reasonable upper limits to the thermal broadening of $\approx$~10~eV, corresponding to $<3000$~km~s$^{-1}$. 
This upper limit provides an upper limit to the Fe temperature of $kT_{\rm Fe}\lesssim 5300$~keV. 
This limit corresponds to a reverse shock velocity of $V_{\rm s} \approx 6900$~km~s$^{-1}$ and the equilibration of the ion temperatures in principle even allows for higher shock velocities.
This velocity is also higher than that inferred from the proper motions of the forward shock. 
But it should be remembered that the shock velocity is a combination of the free expansion velocity of the ejecta and the velocity of the reverse shock in the observed frame: $V_{\rm s}=R/t - dR_{\rm s}/dt$.
If $dR_{\rm s}/dt<0$ --- i.e., the reverse shock moves toward the center, which is reasonably for at those parts of the shell that encountered a high density environment, given that the overall ejecta mass is $\sim 1 M_{\odot}$ \citep[see Figure 2 in][]{Ferreira08} --- the reverse shock velocity may be larger than the forward shock. \par
The large broadening indeed provides an hint for this, but given the uncertainties about the different line-broadening mechanisms, the large line width at the outer strips does not provide sufficient proof for this.
For example, \citet{Sato17b} found both red- and blue-shifted bright knots throughout the remnant covering a velocity range of -8000 to +10000~km~s$^{-1}$. 
Comparing our result with Figure 2 in \citet{Sato17b}, we find some high-velocity compact knots in strips with large line width.

\subsubsection{The ejecta abundances}
As for the elemental abundance of ejecta, we assumed a pure-metal plasma, with the Si abundance set to $10^5$~solar. 
Our fits indicate that the abundance ratio Fe/Si is $\sim 1$, whereas the O/Si abundance is ratio is generally $\sim 0.03$. 
But for some strips we only could obtain an upper limit  (Table \ref{tab:results}).
The lack of the O ejecta is consistent with the previous with {\it Chandra} \citep{Reynolds07}. 
The number ratio of O/Fe is lower than 0.6 at most. 
Such low ratio is cannot be explained by the core-collapsed progenitor and supports that the origin of Kepler's SNR is type Ia SN \citep{Iwamoto99}.

\subsection{The electron temperature and ionization timescale}

Figure \ref{fig:plot1} (c) and (d) show the electron temperature parameter $kT_{\rm e}$ and ionization parameter $n_{\rm e}t$. 
The results for the CSM show that the northern bright rim generally has low-$kT_{\rm e}$ and high-$n_{\rm e}t$ and the southern region has high-$kT_{\rm e}$ and low-$n_{\rm e}t$. 
\citet{Sun19} also found that $kT_{\rm e}$ and $n_{\rm e}t$ show anti-correlation in Kepler's SNR by tessellated mapping of {\it Chandra} ACIS data \citep[see Figure 7 in][]{Sun19}. 
They interpreted such high-$kT_{\rm e}$ and low-$n_{\rm e}t$ regions as newly shocked CSM, and low-$kT_{\rm e}$ and high-$n_{\rm e}t$ for the Si- or S-rich ejecta. 
Although such features might be real, the coupling effect of fitting parameter of $kT_{\rm e}$ and $n_{\rm e}t$ could also lead such results. 
We found the results for the ejecta have the inverse feature to the CSM. 
We should discuss this by using wide-band spectra with 2-Dimensional spatial resolution in future. \par
As a whole, the range of $n_{\rm e}t$ values ($2.5\times 10^{10}$--$13\times 10^{10}$~cm$^{-3}$~s) for the CSM component suggests typical densities of $n_{\rm e} \approx 2$--$10$~cm$^{-3}$, assuming an age of $t=416$~yr, which is the maximum age of the plasma given the age of the SNR. 
In reality, $t$ is likely to be smaller, as much of the plasma will have been heated more recently. 
So the electron densities are likely larger. 
Note, however, that this estimate may be skewed toward the denser parts of the plasma, which have higher emissivities. 
The $n_{\rm e}t$ is smallest in the southern region, consistent with the fainter thermal X-ray emission from that region, as well as the larger shock velocities, which implies less deceleration and lower densities. \par
The variation in electron temperature is relatively limited: $kT_{\rm e}\approx 0.5$--$0.9$~keV. 
This is inconsistent with the measured shock velocities of $V_{\rm s}\approx 2000$--$4500$~km~s$^{-1}$ assuming complete electron-ion equilibration, which would lead to $kT_{\rm e}\approx 5$--$25$~keV.
Clearly, the plasma cannot be fully temperature equilibrated.

\section{Conclusions} \label{sec:conclusion}
In this paper, we reveal the line-of-sight kinematics of the CSM in Kepler's SNR, using a new long observation by the RGS detectors on board the {\it XMM-Newton} satellite using the cross-dispersion direction to spatially resolve the remnant in 1D. 
At the edge of the remnant, we find the CSM is blue-shifted with a velocity of 0--500~km s${}^{-1}$. 
Our result is consistent with a ``runaway'' AGB star as the progenitor of the remnant moving in 3D towards the northwest and in our direction along the line-of-sight \citep{Bandiera87,Borkowski92}. 
In addition, we also derive information on the kinematics at the central bar, and find that the northwestern half is blue-shifted and southeastern half is red-shifted. 
This is consistent with the optical observation of broad H$\alpha$ line \citep{Blair91} and the torus-like shape of CSM distribution \citep{Burkey13}. 
These results are explainable by both the SD and CD scenarios of the progenitor system. 
Future simulations considering the observed line-of-sight velocity in the X-ray band and observation by calorimeters with 2D spatial resolution could help further discussion about the kinematics of CSM around SNRs.

\acknowledgments
We thank the anonymous referee for their comments to improve the manuscript. 
In addition, TK thanks Sbarcea Oriana, Chie Sakuta, Jirina Salkova, Ping Zhou and Lei Sun representing the API SNR group members, Hiroki Akamatsu, Shin'ichiro Ando, and David Gardenier for their kind supports for his short-staying at de Universiteit van Amsterdam. 
TK is supported by the Advanced Leading Graduate Course for Photon Science (ALPS) in the University of Tokyo. 
JV is partially supported by funding from the European Union’s Horizon 2020 research and innovation programme under grant agreement No 101004131 (SHARP).
SK is partly supported by Leading Initiative for Excellent Young Researchers, MEXT, Japan.
JPH acknowledges support for X-ray studies of SNRs from NASA grant number NNX15AK71G to Rutgers University.
This work is parially supported by the Japan Society for the Promotion of Science (JSPS) KAKENHI Grant No. 18H05459, 19K03908, 19K03915, 20H00174, and 21H01121, Shiseido Female Researcher Science Grant, and Graduate Research Abroad in Science Program (GRASP) in the University of Tokyo.

\facilities{{\it XMM-Newton} (RGS) \citep{Herder01}, {\it Chandra} (ACIS-S) \citep{Garmire03}}

\software{SAS 18.0.0 \citep{Gabriel04}, CIAO 4.9 \citep{Fruscione06}, Xspec 12.9.1 \citep{Arnaud96}, HEASoft 6.21 \citep{Blackburn95}}

\bibliography{bibliography}{}
\bibliographystyle{aasjournal}

\end{document}